\documentclass[runningheads]{llncs}

\usepackage{graphicx}
\usepackage{color}
\usepackage{times}
\usepackage{epsfig}
\usepackage{amsmath}
\usepackage{bbm}
\usepackage{amssymb}
\usepackage{multirow}
\usepackage{bm}
\usepackage{soul}
\usepackage{hyperref}

\begin{document}


\title{A Multi-attribute Controllable Generative Model for Histopathology Image Synthesis}



\author{Jiarong Ye\inst{1}\thanks{These authors contributed equally to this work.}, Yuan Xue\inst{1,2}\textsuperscript{*}, Peter Liu\inst{3} \\ Richard Zaino\inst{4}, Keith C. Cheng\inst{4}, Xiaolei Huang\inst{1}}
\institute{College of Information Sciences and Technology, The Pennsylvania State University, University Park, PA, USA
\and Department of Electrical and Computer Engineering, Johns Hopkins University, \\Baltimore, MD, USA
\and Independent Researcher, Upper Dublin High School, Fort Washington, PA, USA
\and College of Medicine, The Pennsylvania State University, Hershey, PA, USA}
\authorrunning{Jiarong Ye~\textit{et al.}}

\maketitle


\begin{abstract}

Generative models have been applied in the medical imaging domain for various image recognition and synthesis tasks. However, a more controllable and interpretable image synthesis model is still lacking yet necessary for important applications such as assisting in medical training. In this work, we leverage the efficient self-attention and contrastive learning modules and build upon state-of-the-art generative adversarial networks (GANs) to achieve an attribute-aware image synthesis model, termed AttributeGAN, which can generate high-quality histopathology images based on multi-attribute inputs. In comparison to existing single-attribute conditional generative models, our proposed model better reflects input attributes and enables smoother interpolation among attribute values. We conduct experiments on a histopathology dataset containing stained H\&E images of urothelial carcinoma and demonstrate the effectiveness of our proposed model via comprehensive quantitative and qualitative comparisons with state-of-the-art models as well as different variants of our model. Code is available at \href{https://github.com/karenyyy/MICCAI2021_AttributeGAN}{\color{blue}{https://github.com/karenyyy/MICCAI2021\_AttributeGAN}}.

\end{abstract}


\section{Introduction}
Discriminative models, especially those based on deep learning, have been proven effective in various medical image analysis tasks~\cite{shen2017deep}. However, such models primarily focus on discovering distinguishable patterns and features existing in medical images for down-stream analysis tasks, thus may neglect patterns that are characteristic of the images but not distinct enough for discriminative tasks. 
Meanwhile, generative models provide a complementary way of learning all image patterns by modeling the entire data distribution. Towards a better comprehension of medical image attributes, we propose an attribute-guided generative adversarial network, termed AttributeGAN, to model the data distribution conditioned on different attributes and link the attribute values with image patterns and characteristics. Different from existing generative models proposed in the medical image domain for applications such as cross-modality translation~\cite{armanious2020medgan}, synthetic augmentation~\cite{xue2021selective} and image reconstruction~\cite{quan2018compressed}, we investigate the problem of synthesizing histopathology images conditioned on different image attributes to build a more controllable and interpretable medical image generative model. 

Existing literature on controllable and interpretable image synthesis models~\cite{shen2020interpreting,shoshan2021gan} focus on noticeable attributes such as human body pose, hair color, age of human face, among others. However, attributes of medical images are more nuanced and harder to model and thus the problem of generating medical images based on controllable attributes is more challenging to solve. For conditional image synthesis, conditional GANs (cGANs)~\cite{odena2017conditional,miyato2018cgans} have utilized various types of discriminator networks to help the models capture the relationships between input conditions and image features. However, few of them work on multiple attribute inputs or are studied for medical image applications.

In this work, our goal is to develop an attribute-guided medical image synthesis model which can generate high-resolution and realistic images as well as make sure the generated images accurately reflect the attributes given to the model. We build upon a successful unsupervised generative model, leverage a carefully designed attribute-attention model, and employ a conditional contrastive learning strategy to efficiently model the conditional data distribution. Multiple attributes are one-hot encoded and concatenated with the noise vector and fed into different stages of the proposed model. Our proposed model generates photo-realistic histopathology images while being more controllable and interpretable than unconditional generative models. We conduct experiments on a histopathology dataset containing stained H\&E images of urothelial carcinoma and compare our proposed AttributeGAN with the state-of-the-art cGAN as well as different variants of our model. We summarize our contributions in this work as follows:
\begin{itemize}
    \item[*] We propose a multi-attribute controllable generative model for high quality histopathology image synthesis. To the best of our knowledge, our work is the first to develop an attribute-aware GAN model with the capability to precisely control cellular features while preserving photo-realism for synthesized images.   
    \item[*] We incorporate efficient attention modules and conditional contrastive learning in both the generator and the discriminator to significantly improve quality as well as achieve better attribute-awareness of the generated images. Experiments on a histopathology dataset show better image quality using our proposed AttributeGAN than the state-of-the-art conditional GAN model.
\end{itemize}

\section{Methodology}


\begin{figure}
\includegraphics[width=\textwidth]{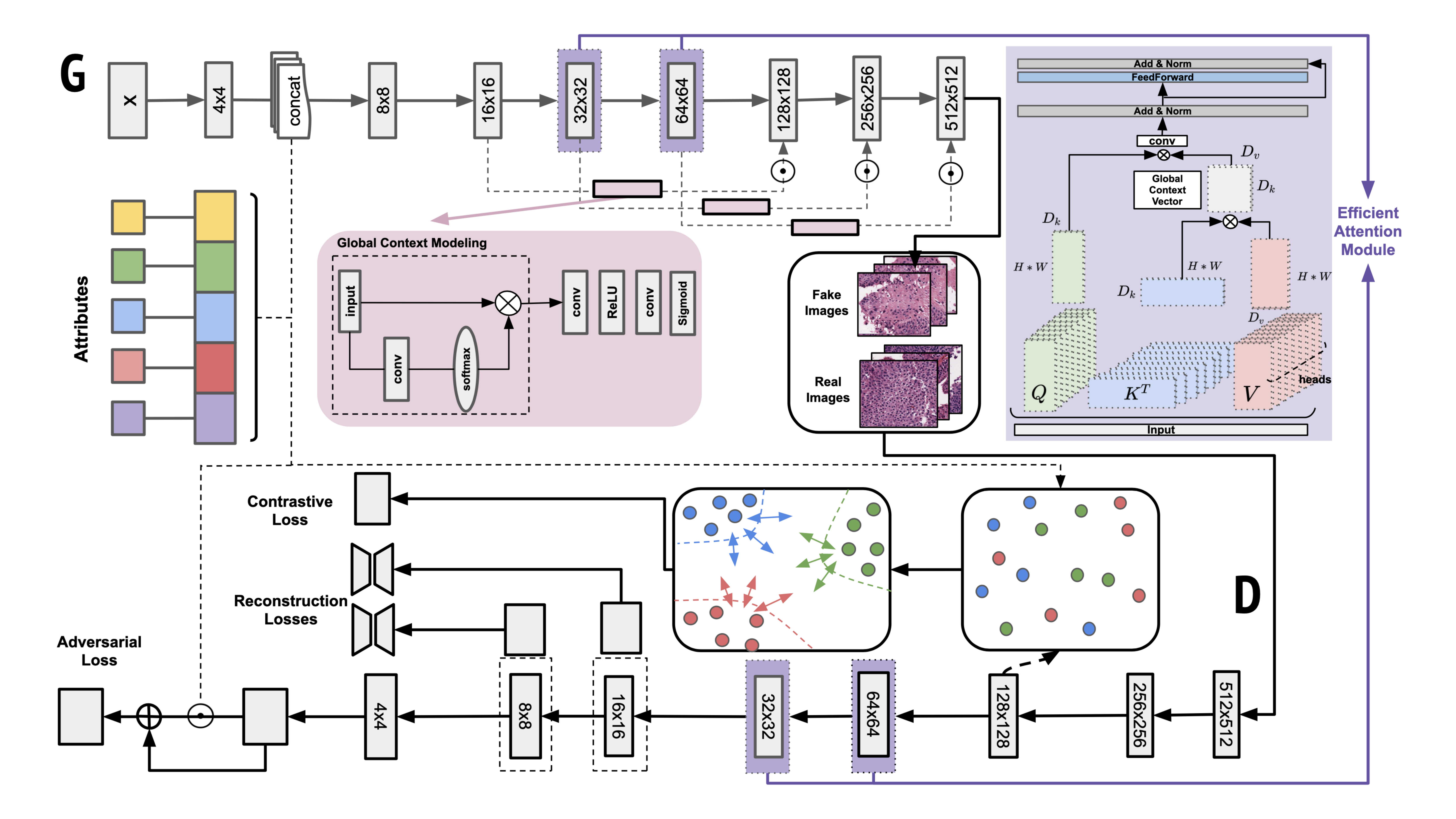}
\caption{The architecture of our proposed controllable cellular attribute-aware generative model. Each color block of the attribute vector input represents a corresponding cellular attribute feature: cell crowding, cell polarity, mitosis, prominence of nucleoli and state of nuclear pleomorphism. Different colors in the feature space for contrastive learning refers to the label constructed by combination of 5 cellular attribute levels (\textit{e.g. (cell-crowding-severe, cell-polarity-completely-lacking, mitosis-frequent, nucleoli-prominent, pleomorphism-moderate)})} \label{architecture}
\end{figure}

To guarantee the quality of synthesized images, we build our model upon a recent unsupervised backbone generative model introduced by \cite{Liu2021TowardsFA}. For attribute-aware and controllable generation, we incorporate multi-attribute annotations of each image as the additional condition information to explicitly control the generation process. With attribute conditions inserted, the synthesized results are expected to 
maintain sufficiently photo-realistic while accurately capturing the distinguishable image feature patterns within attributes. To fulfill the first goal, we adopt a skip-layer channel-wise excitation (SLE) module and include additional reconstruction loss in discriminator as in \cite{Liu2021TowardsFA}. SLE leverages learned feature patterns from a lower abstract level to further re-calibrate the channel-wise features map of higher scale. As demonstrated in the architecture of our proposed controllable cellular attribute-aware generative model in Fig.~\ref{architecture}, in addition to the basic structure of SLE, we further improve the backbone by incorporating a global attention pooling for context modeling~\cite{Cao2020GlobalCN} at earlier stages of upsampling before the transformation through the bottleneck blocks to capture channel-wise dependencies.

For the second goal of attribute learning, while existing conditional GANs (cGANs)~\cite{miyato2018cgans} concatenate noise vectors with the conditional vectors and leverage projection discriminator for condition correctness, such models may not be able to capture the nuanced changes in attribute levels of medical images. 
In addition to input concatenation and projection discriminator, we integrate conditional contrastive losses~\cite{kang2020contragan} to both discriminator and generator to exploit the relation between images and the attributes they contain. 
Integrating a self-supervised learning based module to exploit data-to-data and data-to-attribute relations within a mini-batch of proper size comes with two merits. First, with known attributes available for reference, the performance no longer heavily relies on the hard negative and positive samples mining. We consider the anchor image itself together with real images with the same attribute combination as positive samples, while real images with different attribute combinations in the same mini-batch as negative samples. Second, the performance of capturing the explicitly distinguishable feature representations in a fine-grained manner is substantially enhanced. Theoretically, 
this is achieved by minimizing the feature-level distances between positive samples while maximizing the distances between positive and negative samples. During training, the knowledge of attribute-dependent feature distinction learned by the discriminator is then passed to the generator for synthesizing images that are more sensitive to inter/intra-attribute characteristics. The effectiveness is further proven empirically in the qualitative ablation study of model architecture with and without the contrastive learning objective as demonstrated in Fig.~\ref{ablation}.  
To elaborate, first we denote $X = \{x_1, x_2, ..., x_n\}$ as extracted features from the intermediate stage of discriminator, and $Y = \{y_1, y_2, ..., y_n\}$ as the combination of multiple attributes. Intuitively, after mapping data to the hypersphere via feature and attribute projectors $f$ and $l$, our goal is to push the inter-attribute samples further and pull intra-attribute ones closer at the feature level. Thus, our conditional contrastive loss is formulated as:

\begin{equation}
\mathcal{L}(x_i, y_i; t) = - \log \left ( \frac{
\exp(\frac{f(x_i)^\top l(y_i)}{t}) + 
\sum^n_{k=1} \mathbbm{1}_{y_k = y_i} \cdot \exp(\frac{f(x_i)^\top f(x_k)}{t})
}{
\exp(\frac{f(x_i)^\top l(y_i)}{t}) + 
\sum^n_{k=1} \mathbbm{1}_{k \neq i} \cdot \exp(\frac{f(x_i)^\top f(x_k)}{t})
} \right ),
\end{equation}
where $\mathbbm{1}$ is the indicator function and the scalar value $t$ plays the role as the regularizer to balance the push and pull force among samples across different and within the same group of attributes.

Recent GAN models~\cite{zhang2019self} for image synthesis have adopted the self-attention module~\cite{vaswani2017attention} to capture long-range dependencies within the image. However, the dot-product based self-attention can quadratically increase computational complexity and constrain the number of images inside each batch. Meanwhile, the aforementioned contrastive learning efficiency heavily relies on a relatively large batch size as both data-to-data and data-to-attribute relation learning would be seriously compromised with a small batch size and insufficient number of positive/negative pairs. Hence, in order to free up more space to accommodate a larger volume of data in each batch and train with lower computational complexity, we apply a more efficient equivalence~\cite{shen2021efficient} of self-attention. As illustrated in the efficient attention module in Fig.~\ref{architecture}, feature vectors at intermediate stages in both generator and discriminator are projected onto three latent spaces through convolution operations termed as query, key and value as in the original self-attention~\cite{vaswani2017attention}, and denoted as $Q, K, V$, respectively. Here, $Q \in \mathbb{R}^{(H*W) \times d_q}, K \in \mathbb{R}^{(H*W) \times d_k}, V \in \mathbb{R}^{(H*W) \times d_v}$, $d_q = d_k$ and $H, W$ refer to the height and width of the image. Leveraging the associative property of matrix multiplication, rather than start with the multiplication of $QK^T$ as formulated in \cite{vaswani2017attention} to measure the pair-wise similarity exhaustively, instead we begin with the multiplication between $K^T$ and $V$. It is feasible because $(\frac{QK^T}{n})V = \frac{Q}{\sqrt{n}}(\frac{K^T}{\sqrt{n}}V)$.
Following this procedure, we obtain a matrix $g \in \mathbb{R}^{d_k \times d_v}$, representing the intermediate global context vector with dimension of $d_v$ in $d_k$ channels after aggregating from $H*W$ positions through weighted summation. At the next step, the context vector is acquired by having each pixel gathering positional features from all $d_k$ channels for $d_v$ dimensions, by multiplying $Q$ and the result of $(K^TV)$. With the efficient attention, the memory complexity is reduced to $\mathcal{O}(d_k*d_v)$ from the original $\mathcal{O}(n^2)$, escalating convergence speed and freeing up more space, making it possible for conditional contrastive learning to deliver its performance to the fullest.

More specifically, for conditional attributes, we encode the input condition into a one-hot vector with attribute level labels for all five cellular features. The attribute vector is later concatenated with the input noise vector after the initial stage of upsampling both vectors using transposed convolution operators. For synthesizing images with resolution $512 \times 512$, efficient attention modules are applied in two intermediate upsampling stages at $32 \times 32$ and $64 \times 64$ resolutions as shown in Fig.~\ref{architecture}. For each upsampling block without attention module, input images first go through an upsampling layer with scale factor set as 2, immediately followed by a gaussian blurring kernel for antialiasing. Next, to enlarge the feature learning space channel-wise, a basic block including a convolutional layer, a batch normalization layer and an activation layer is added as another major component in each individual upsampling block. Gated Linear Units (GLU) is utilized for every activation layer in the AttributeGAN architecture, as it has shown quality-improving potential over the commonly used ReLU or GELU activations~\cite{shazeer2020glu}. Additionally, three skip-layer connections are applied at the resolutions of $16\times16$, $32 \times 32$, and $64 \times 64$ to $128 \times 128$, $256\times 256$ and $512 \times 512$ in order to strengthen the gradient signals between layers.

For the discriminator, the conditional attributes are required together with either synthesized or real images to be further utilized in a projection based discrimination. Attribute vectors are fed into a feed-forward layer before being incorporated into the output of discriminator. As shown in Fig.~\ref{architecture}, at the resolution of $128 \times 128$ the feature vectors and the attribute level information are projected to an embedded space for contrastive learning, which is later included in the losses for the discriminator. To further refine the discriminator's capability of capturing a more comprehensive feature map to be differentiated from the fakes, two auxiliary reconstruction losses are added. We utilize two additional simple decoders trained within the discriminator for the $8 \times 8$ and $16\times16$ feature vectors, and calculate the mean squared error (MSE) for both in the reconstruction loss.


\begin{figure}[tbp]
\includegraphics[width=0.96\textwidth]{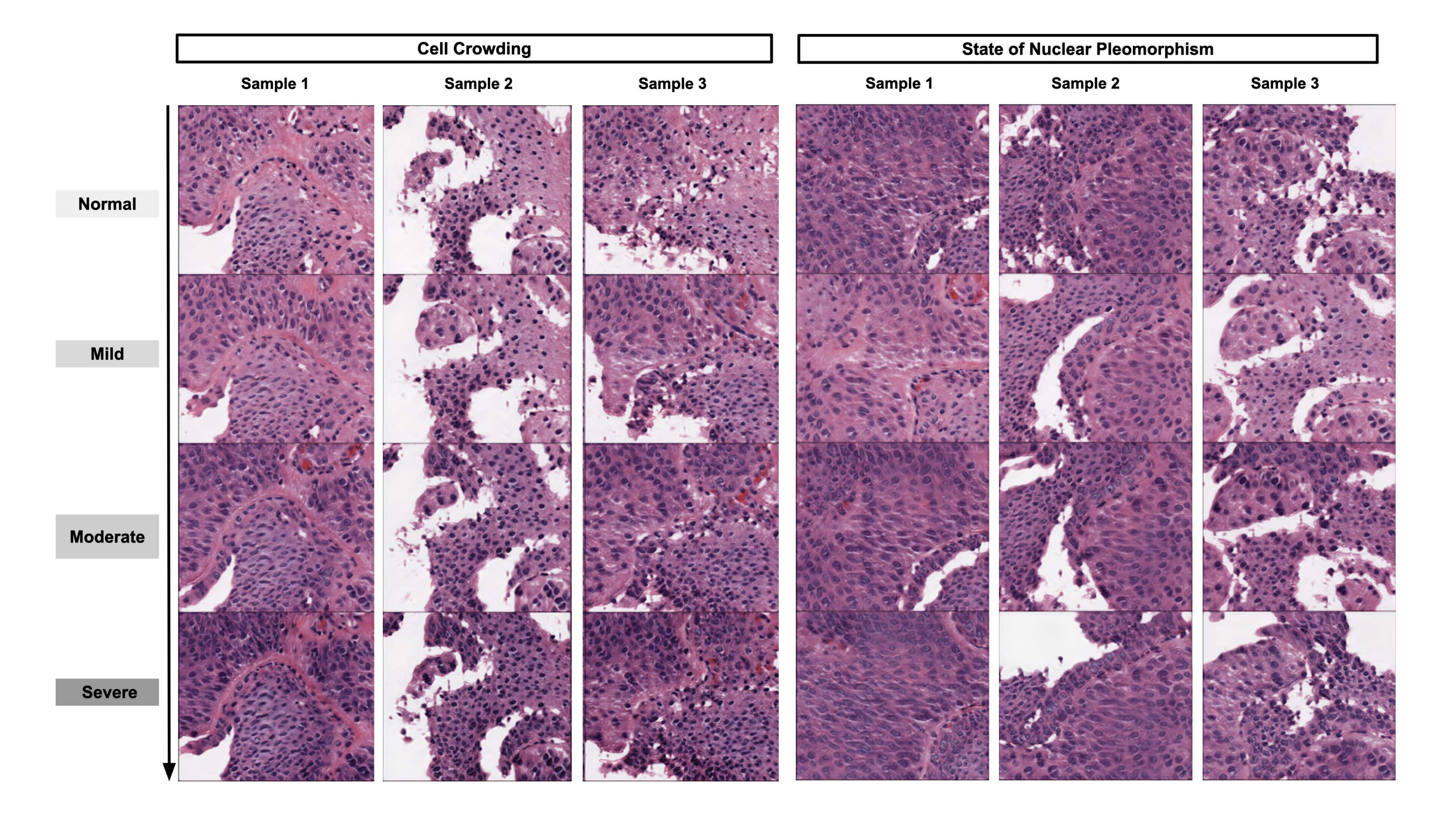}
\caption{The AttributeGAN generated histopathology patches based on different levels of cell crowding and the state of nuclear pleomorphism. The input noise vector is set to be identical for each sample, which explains the resemblance of general shape and texture shared within each column and rich diversity across columns. Zoom in for better view.} 
\label{results}
\end{figure}

\section{Experiments and Results}

\noindent \textbf{Dataset}. We conduct comprehensive experiments on a histopathology dataset representing patients with bladder cancer collected by \cite{zhang2019pathologist}. The dataset contains $4,253$ histopathology image patches with $512\times 512$ resolution. Each patch is accompanied with a paragraph of pathology report descriptions provided by multiple experienced pathologists.
Each report follows a template format that describes 5 types of key morphological visual cellular features essential for classifying urothelial carcinoma, including cell crowding, cell polarity, mitosis, prominence of nucleoli and state of nuclear pleomorphism.

To achieve a more concise representation of attributes and their levels, we extract feature-describing keywords in the report as annotations (see Table 1-5 in Supplementary Materials).
Converting raw reports to categorical levels for each cellular attribute facilitates the manipulation of semantic editing in our experiments, as demonstrated in Fig.~\ref{results}. There are 4, 3, 3, 2, 4 levels assigned to describe different degrees of cell crowding, cell polarity, mitosis, nucleoli and pleomorphism, respectively. Following this procedure, each patch is paired with a combination of levels from all 5 cellular attributes. To accelerate the learning of attribute-relevant patterns, we discard the combinations with frequency less than the $20^{th}$ percentile since most of those merely appear in the dataset once or twice. 

\noindent \textbf{Implementation Details}. 
AttributeGAN is trained on two NVIDIA Quadro RTX 6000 GPUs each with 24GB RAM in parallel by applying the PyTorch DistributedDataParallel module together with SyncBatchNorm. The GPU space freed up from the attention module efficiency enables a larger batch size. In our experiments, the batch size is set as 64, and each device processes half of the inputs from the current batch. The learning rate is fixed to be $2e-4$ throughout the entire 50000 steps of training.


\begin{figure}[tbp]
\includegraphics[width=0.9\textwidth]{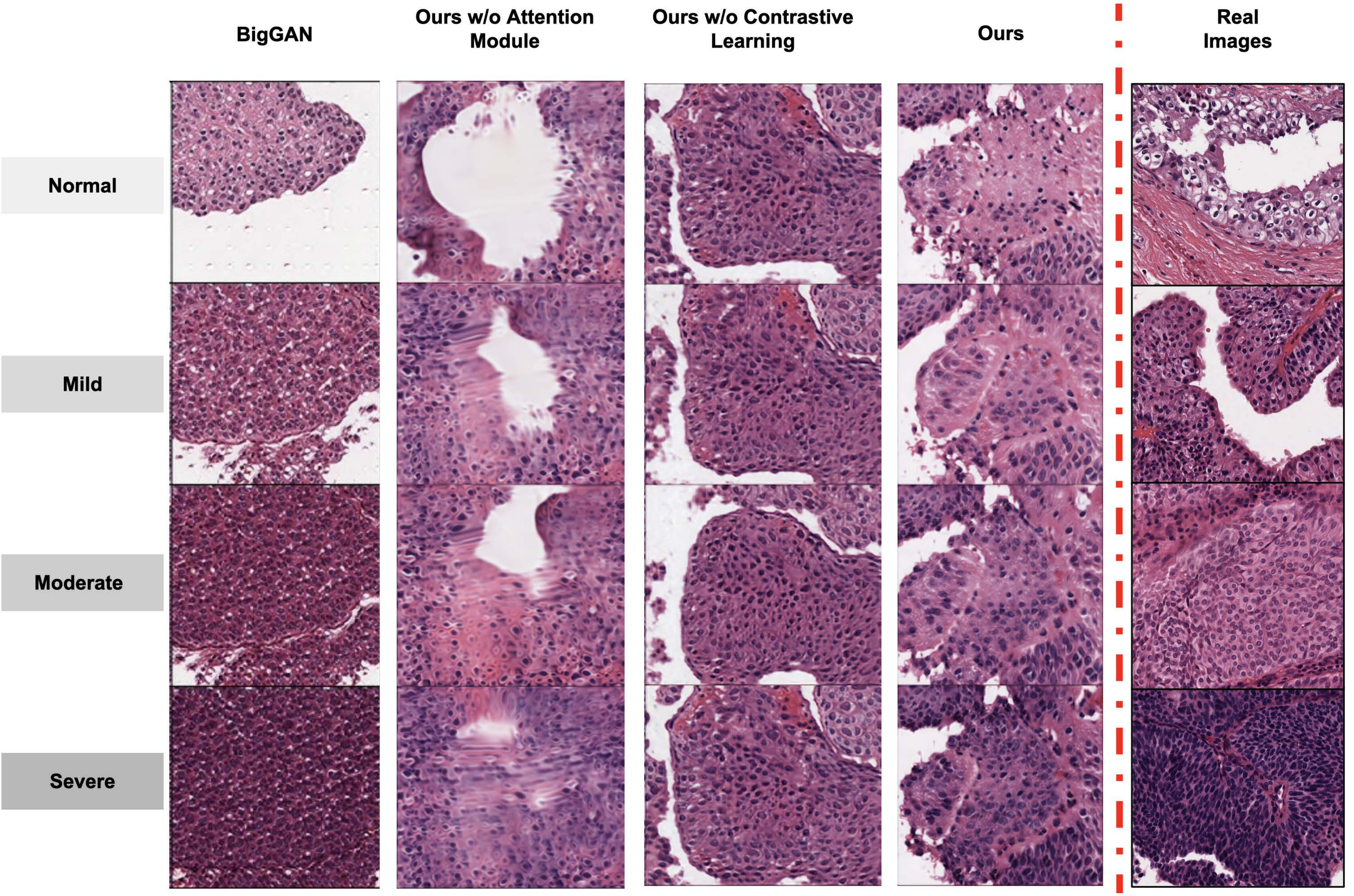}
\caption{Generated images from ablation study. The selected attribute for ablation study is cell crowding. We present the comparison among images synthesized using our baseline model BigGAN, our proposed AttributeGAN w/o the efficient attention module, our proposed AttributeGAN w/o the conditional contrastive loss, and real images for 4 levels of cell crowding: normal, mild, moderate and severe. Zoom in for better view.} 
\label{ablation}
\end{figure}

We present example images generated by our AttributeGAN in Fig.~\ref{results}. To demonstrate the smooth progression through different attribute levels and showcase disentanglement among attributes, the input attribute is framed as a 5-dimensional vector where we only alter one attribute at a time inside each result batch. Other than the attribute whose level is being varied, the remaining dimensions are fixed to be a combination of the other four attributes that frequently appear in the dataset. With attribute conditions given in such manner, the generated images show clear progressions in cellular pattern in accordance with the changes in input attribute condition. 

To examine the effectiveness of our proposed AttributeGAN and its components, we compare images generated by different models as well as the real images in Fig.~\ref{ablation}. 
Various well-developed and extensively-used models are relevant to attribute-controlling, such as Fader Networks \cite{lample2017fader}, StyleGAN \cite{karras2019style} and StyleGAN v2 \cite{karras2020analyzing}. Although the aforementioned models present state-of-the-art results on photo-realism and attribute accuracy, they are not suitable to be directly compared with our approach for the conditional histopathology image synthesis task, because they are designed for slightly different goals such as semantic editing of assigned attributes (e.g. the Fader Networks), or unconditional image synthesis (e.g. StyleGAN, StyleGAN v2). 
Hence we consider the state-of-the-art conditional GAN model, BigGAN~\cite{brock2018large}, as the most appropriate baseline model. Since BigGAN can only handle single-dimensional condition, we train 5 different BigGAN models for different attributes. Considering that BigGAN consumes larger memory and requires longer time to converge, we train all baseline BigGAN models with image resolution $256 \times 256$. During comparison, we resize all images to the same size for fair comparison.  One can observe that images generated by our models show superb realism. Compared with the BigGAN model, different variants of AttributeGAN model keep the global shape and texture well inside each column. On the contrary, the global image pattern changes for BigGAN given different attribute level inputs. For variants of our AttributeGAN, our proposed model without the attention module generates less realistic images, and the model without the conditional contrastive learning reacts less responsively to the changes in attribute level. The full AttributeGAN model respects the changes in attribute level and retains the global patterns well. 

\begin{table*}[tbp]
\begin{center}
\begin{tabular}{l|c|c|c|c|c|c}
    \hline
    \multirow{3}{*}{\textbf{Methods}} &
      \multirow{3}{*}{\textbf{FID}$\downarrow$}  &
      \multicolumn{5}{c}{\textbf{Attribute Error}$\downarrow$} \\
    \cline{3-7}  
    {}&{}& Cell & Cell & \multirow{2}{*}{Mitosis} & \multirow{2}{*}{Nucleoli} & \multirow{2}{*}{Pleomorphism} \\
      {}&{}& Crowding & Polarity & {} & {} & {}  \\
    \hline
    Real Images* & - & .011 & .034 & .037 & .018 & .014 \\
    \hline
    BigGAN \cite{brock2018large} & 158.39 & .112 & .080 & .104 & \textbf{.049} & .065 \\
    \hline
    AttributeGAN (Ours) & \multirow{2}{*}{142.015} &  \multirow{2}{*}{.035} & \multirow{2}{*}{\textbf{.078}} & \multirow{2}{*}{.208} & \multirow{2}{*}{.056} & \multirow{2}{*}{\textbf{.023}} \\
    \enspace \textit{w$\slash$o} EA  &  &  &  &  &  &  \\
    \hline
     AttributeGAN (Ours) & \multirow{2}{*}{55.772} &  \multirow{2}{*}{.094} & \multirow{2}{*}{.112} & \multirow{2}{*}{.111} & \multirow{2}{*}{.056} & \multirow{2}{*}{.070} \\
    \enspace \textit{w$\slash$o} CCL   &  &  &  &  &  &  \\
    \hline
     AttributeGAN (Ours) & \textbf{53.689} & \textbf{.021} & .098 & \textbf{.088} & .081 & .063 \\
    \hline
  \end{tabular}
  \end{center}
  \caption{Quantitative evaluation results of different methods.
Real images are from a holdout validation set during fine-tuning of the pre-trained classifier. Note that we report BigGAN results from five independently trained BigGAN models for five attributes as it can only work with single attribute inputs. EA refers to Efficient Attention module and CCL refers to the Conditional Contrastive Loss.}\label{result_table}
\end{table*}

In Table~\ref{result_table}, we show quantitative comparison results between different models. Following conventions in image synthesis works~\cite{zhang2019self,brock2018large}, we adopt Fréchet Inception Distance (FID)~\cite{heusel2017gans} score which has shown to correlate well with human perception of realism. FID measures the Fréchet distance between two multivariate Gaussians fit to features of generated and real images extracted by the pre-trained Inception V3~\cite{szegedy2016rethinking} model. Compared with BigGAN whose FID score is averaged from five BigGAN models, all AttributeGAN variants achieve better FID score indicating better realism. After including the attention module, the FID score improved significantly for the AttributeGAN model. To better evaluate the correctness of represented attributes, we further calculate an Attribute Error to measure the discrepancy between attribute levels predicted by an ImageNet pre-trained ResNet18~\cite{he2016deep} model fine-tuned on the histopathology dataset and the groundtruth attribute levels. Images generated by all models are first normalized to same resolution $224 \times 224$ for fair comparison. All attribute levels are normalized to the range $[0,1]$ and the MSE of the predicted attributes and the groundtruth attributes are computed as the attribute error value. During fine-tuning of the ResNet18 model, we keep a holdout validation set and the corresponding attribute error evaluated on the holdout real images are also reported in Table~\ref{result_table}. For BigGAN and our proposed AttributeGAN without attention, although they achieve small attribute errors for certain attributes, the quality of generated images are lower which makes them differ more from real images, thus the attribute prediction model trained on real images may not be able to correctly predict the attribute level for such images. Compared to AttributeGAN without contrastive learning, the full AttributeGAN generally gets lower attribute error, especially on cell crowding. Based on both the qualitative and quantitative comparisons, we prove the necessity of the attention module and the conditional contrastive loss, and show that one multi-attribute AttributeGAN model can generate images with better quality than multiple BigGAN models for conditional histopathology image synthesis.

\section{Discussion}

To assess the quality of the generated images and how well the images correspond to the input attribute levels, we presented five sets of images that were generated based on different cellular attribute levels to two expert pathologists.  Both pathologists commented that the synthetic images are remarkably good in resembling routinely stained H\&E images of urothelial carcinoma.  In the set of images generated according to different levels of cell crowding (see examples in Fig.~\ref{results}-Left), the crowding of nuclei occurs appropriately overall at each of the described levels, and the degree of crowding remains within the realm of reality, although for a few images, the increase in crowding seems to be by increasing the epithelial/stromal ratio, rather than increasing the density of cells within the same amount of epithelium.  For the set of images generated according to different levels of pleomorphism (see examples in Fig.~\ref{results}-Right), an increase in nuclear pleomorphism was observed as the images progress through the pleomorphism prominence levels. For the other three sets of images generated based on different levels of cell polarity, mitosis, and prominence of nucleoli (see Figures 1-3 in Supplementary Materials), the pathologists commented that no obvious progression was observed through those sequences of images.  We plan to further investigate these three attributes in our future work, study whether the attributes are correlated in some fashion in real images and learn how to improve the responsiveness of generated images to varying input conditions.

\section{Conclusion}
In this work, we present a multi-attribute guided generative model, AttributeGAN, for synthesizing highly realistic histopathology images. Images generated by the proposed model show smooth progression through different input attribute levels and contain photo-realistic patterns. With the quality of synthesized images, AttributeGAN can be potentially used for medical education or training and support various medical imaging applications.



\bibliographystyle{splncs04}
\bibliography{ref}

\end{document}